\documentclass[aps,twocolumn,groupedaddress,floatfix,showpacs]{revtex4} 

\usepackage{epsfig}
\usepackage{amssymb}
\usepackage{amsmath}

\begin{document}

\title{Low-temperature quantum fluctuations in overdamped ratchets}

\author{Stefan A. Maier\footnote{Permanent address: Institute for Theoretical Physics C, RWTH Aachen University, 52062 Aachen, Germany} %
 and Joachim Ankerhold}
\affiliation{Institut f\"ur Theoretische Physik, Universit\"at Ulm, Albert-Einstein-Allee 11, 89069 Ulm, Germany}

\date{\today}

\begin{abstract}
At low temperatures and strong friction the time evolution of the density distribution in position follows a quantum Smoluchowski equation. Recently, also higher-order contributions of quantum fluctuations to drift and diffusion coefficients have been systematically derived.
As a non-trivial situation to reveal the impact of subleading quantum corrections and to demonstrate convergence properties of the perturbation series, directed transport in ratchets is studied.  It is shown that the perturbation series typically has a non-monotonous behavior. Depending on symmetry properties higher-order contributions may even compensate current reversals induced by leading quantum fluctuations. This analysis demonstrates how to consistently treat the dynamics of overdamped quantum systems at low temperatures also in numerical applications.
\end{abstract}


\maketitle

\section{Introduction}
Directed transport in ratchet-type of force fields has attracted substantial research for more than a decade now as one prototype of Brownian motor (see e.g.\ \cite{astumian_2002,reimann_2002,hanggi_2009}). These engines perform work by extracting energy from out-of-equilibrium fluctuations of the environment. Realizations can be found in a variety of contexts ranging from biological systems to designed mesoscopic devices such as electrical circuits containing Josephson junctions. The corresponding classical theory is well established and based on Langevin equations and, equivalently, on Fokker-Planck equations for the phase-space dynamics.

Much less is known for the corresponding quantum mechanical problem, mainly due to the fact that in this regime the interplay of quantum fluctuations in the surrounding heat bath and system dynamics generates long-range retardation effects in time so that a simple time evolution equation for the density distribution does not exist. Progress has been made particularly in the adiabatic limit for a dichotomous noisy drive  and for tight-binding systems \cite{reimann_1997,goychuk_1998,grifoni_2002,grifoni_2006,muehlbacher_2009}. A considerable simplification occurs in the strong-friction range. Then, the time scale for relaxation in position by far exceeds all other time scales and a time-local evolution equation for the marginal position distribution can be derived from the formally exact path-integral expression. This so-called quantum Smoluchowski equation (QSE) \cite{ankerhold_2001} applies also to the low-temperature regime, where the energy scale for friction $\hbar\gamma$ (friction constant $\gamma$) is much larger than the thermal energy scale $k_{\rm B} T$. Conceptually, this domain is far from being classical which requires just the opposite condition $\gamma\hbar\ll k_{\rm B} T$, even though quantum fluctuations can be treated perturbatively. Their leading contributions are of order $\ln(\gamma)/\gamma$ and typically have strong impact in barrier escape problems.

This is particularly true for transport in ratchets as shown in detail in a series of papers \cite{machura_2004,machura_2006,dajka_2007}. Net currents are extremely sensitive to variations of the potential topology so that weak quantum fluctuation may e.g.\ even generate current reversals. What has not been done though, is a systematic analysis of higher-order quantum corrections. This is the goal of the present work based on the general perturbative approach for drift and diffusion coefficients in the QSE recently developed in \cite{ours_2010}.
The naive expectation is that subleading quantum fluctuations lead to monotonous convergence properties. In contrast, our results indicate that typically this is not the case and even current reversals may be completely compensated by higher-order contributions in the perturbation series. It turns out that a consistent treatment has to take into account orders in perturbation theory, which carry information about non-local characteristics of the ratchet potential. More generally, the results of this study show how subleading quantum fluctuations must be consistently incorporated in an extended QSE to describe the real-time dynamics of strongly condensed phase systems at very low temperatures.

\section{Quantum Smoluchowski equation}\label{sec:QSE}

In preceding publications \cite{ankerhold_2001,ours_2010}, we have considered quantum dissipative systems in the limit of strong friction, in which a typical damping constant $ \gamma $
by far exceeds a characteristic frequency $ \omega $ of the isolated system. In the corresponding classical domain, the so-called Smoluchowski regime is characterized by a separation of time scales, whereas the momentum decays on the time scale $ 1/\gamma $, while its position relaxes on a much larger scale $ \gamma / \omega^2 $
 \cite{skinner_1979,brinkman_1956,risken}. The relevant dynamics of this latter process is described by the Smoluchowski equation (SE) \cite{smoluchowski_1916}
\begin{equation*}
 \frac{\partial}{\partial t} P(q,t) = \frac{1}{m \gamma} \frac{\partial}{\partial q} \left[ V'(q) + \frac{\partial}{\partial q} \frac{1}{\beta} \right] P(q,t)
\end{equation*}
for the marginal position distribution $ P(q,t) $ of a particle with mass $ m $ in a potential $ V (q) $  surrounded by an environment at inverse temperature $ \beta = 1/(k_{\rm B} T)$.\\

Quantum mechanically the situation is much more involved. The intrinsic thermal time-scale for fluctuations of the bath $\hbar\beta$ comes into play such that even an environment with white-noise characteristics in the classical domain exhibits colored-noise fluctuations at lower temperatures. Accordingly, simple dynamical equations for the reduced density matrix (results from the full density matrix of system+reservoir after tracing out environmental degrees of freedom) do not exist. For strong friction, an effectively Markovian diffusion-equation for the diagonal part of the density, the position distribution $P(q,t)$, can thus only be derived if the time-scale separation is extended to include also the thermal time-scale. The corresponding QSE has been obtained in leading order in the quantum fluctuations in \cite{ankerhold_2001} and since then has been applied to a variety of transport problems (see \cite{ankerhold_qt_semicl} and references therein).
Recently, subleading quantum corrections to this original QSE have been systematically derived  within a type of semiclassical analysis in \cite{ours_2010} to which we refer for further details. Here we only recall the main features. A general QSE can be cast in the form
\begin{align} \notag
 \frac{\partial}{\partial t} P(q,t) & = \frac{\partial}{\partial q} {\cal L} P(q,t)\, ,\\ \label{eqn:num_curr}
 {\cal L} & = d(q) \left[ D_1 (q) + \frac{\partial}{\partial q} D_2(q) \right]
\end{align}
with drift and diffusion coefficients $D_1$ and $D_2$, respectively. In principle, these coefficients must be derived from the full path-integral dynamics in real time. However, in case of a system with time-independent potential $V(q)$, it turns out that they can be determined by thermal-equilibrium properties. The corresponding Euclidian path-integral approach shows that in the strong-friction regime the thermal distribution in position $P_\beta (q)$ is obtained as
\begin{equation}
\label{thermal}
 P_\beta (q) = Z^{-1} F(q) \, {\rm e}^{- \psi({q}) / \hbar}\, ,
\end{equation}
with a minimal Euclidian action $\psi(q)$, a fluctuation pre-factor $F(q)$, and a normalization $Z$. Then, drift and diffusion coefficients in (\ref{eqn:num_curr}) can be read off from these quantities as
\begin{align*}
 D_1(q)  = \frac{1}{\hbar \beta F(q) }\, \frac{d\psi(q)}{d q}\, , \quad
 D_2 (q)  = \frac{1}{\beta F(q)}\, .
\end{align*}
Note that this form of the coefficients guarantees that spurious equilibrium currents are avoided in each order of perturbation theory \cite{machura_2004,ours_2010}.
 The factor $d(q)$ in (\ref{eqn:num_curr}) captures dynamical and additional thermal corrections and is obtained from the full time-evolution of the reduced density matrix.
It turns out that up to the order in perturbation theory, where analytical results are available, dynamical corrections are completely classical and are at most of order $\omega^2/\gamma$, while additional thermal fluctuations are fixed through the exactly solvable case of a harmonic system \cite{pechukas_2000}. We note in passing that
the current operator in (\ref{eqn:num_curr}) may also be cast in the standard Fokker-Planck form $\tilde{D}_1+\partial_q \tilde{D}_2$ with $\tilde{D}_1=d D_1-(\partial_q d) D_2$ and $\tilde{D}_2=d D_2$.

As will be seen in detail below, quantum corrections in the coefficients of the QSE carry higher than first order derivatives of the potential $V(q)$. This is attributed to the general tendency of quantum fluctuations for delocalization in position and in momentum in accordance with the uncertainty principle. For externally driven systems with  time-dependent potentials $V(q,t)$ the semiclassical evaluation of the full path-integral dynamics reveals that the results of the static case are simply generalized by replacing $V(q)\to V(q,t)$ if the external driving happens to be sufficiently smooth and if its typical frequency $\Omega$ respects the time-scale separation $1/\gamma, \hbar\beta, \hbar\beta \ll 1/\Omega$ \cite{dillenschneider_2009,ours_2010}.

To analyze the impact of higher-order quantum corrections in more detail, we concentrate in the sequel on the low-temperature Smoluchowski regime given by $ 1/\gamma \ll \hbar \beta \ll \gamma / \omega^2 $. A systematic perturbative treatment of the thermal distribution (\ref{thermal}) is then based on a power series in the small parameter
\begin{equation*}
 \lambda  = \frac{\hbar}{\pi m \gamma} \left[ \Psi \left( 1 + \frac{\hbar \beta \gamma}{2 \pi} \right) + C_{\rm E} \right]
  \approx \frac{\hbar}{\pi m \gamma} \ln \left(  \frac{\hbar \beta \gamma}{2 \pi} \right) \, ,
\end{equation*}
where $ \Psi $ denotes the digamma function and $ C_{\rm E} $ is the Euler-Mascheroni constant.  In leading order (order $\lambda$) the current operator is found to read
\begin{equation}  \label{eqn:moyals0}
D^{(0)}_1 \approx  V'(q) \, , \quad D^{(0)}_2  \approx  \frac{1/ \beta}{ 1 - \beta  V''(q) \lambda } \, , \quad
 d^{(0)} \approx \frac{1}{m \gamma} \, ,
\end{equation}
which gives the QSE already used previously, see e.g.\ \cite{ankerhold_2001,machura_2004,machura_2006,dajka_2007}.

In next order (order $\lambda^2$), the path integral for the equilibrium distribution $P_\beta(q)$ is solved by taking into account local harmonic properties of the potential around the endpoint $q$ of the closed minimal-action paths.  Hence, we find
\begin{align} \notag
D^{(1)}_1 &\approx  V'(q) \left[ 1 + \frac{\beta^2}{4}{V'' (q) }^2\lambda^2 \right] \, ,\\ \notag
  D^{(1)}_2 & \approx  \frac{1/ \beta}{ 1 - \beta  V''(q) \lambda + \frac{3}{4} \beta^2  V''(q)^2 \lambda^2} \, , \\
\label{eqn:moyals1}
 d^{(1)} &\approx \frac{1}{m \gamma} \left[ 1 - \frac{\beta^2}{4} V'' (q)^2 \lambda^2 \right] \, .
\end{align}
For even higher-order terms, local anharmonicities are considered as well. Analytical results are available up to contributions of order $\lambda^2/L$ with an anharmonicity length-scale $L$, namely,

\begin{align} \notag
D^{(2)}_1 &\approx  V'(q) \left\{ 1 + \frac{\beta^2}{4} \left[ {V'' (q) }^2 - 2 V'(q) V'''(q) \right] \lambda^2 \right\} \, ,\\ \notag
  D^{(2)}_2 & \approx  \frac{1/ \beta}{ 1 - \beta  V''(q) \lambda + \beta^2 \left[ \frac{3}{4} V''(q)^2 + V'(q) V'''(q)  \right]\lambda^2} \, , \\
\label{eqn:moyals2}
 d^{(2)} &\approx \frac{1}{m \gamma} \left[ 1 - \frac{\beta^2}{4} V'' (q)^2 \lambda^2 \right] \, .
\end{align}
The drift and diffusion coefficients specified in (\ref{eqn:moyals1}) and (\ref{eqn:moyals2}), respectively, provide the systematic extension of the leading-order result (\ref{eqn:moyals0}) and are applied in the remainder to reveal their role in diffusion processes in ratchets.

\section{Transport in adiabatically driven ratchets}

We consider a tilted $ L $-periodic potential of the form
\begin{align*}
 V (q) = & V_0 \left\{ \sin ( 2 \pi q/L) +a \sin \left[ 4 \pi ( q/L -b )  \right] \right.\\
        & \left. + c \sin \left[ 6 \pi (q/L-b) \right]\right\} \, ,
\end{align*}
which lacks an inversion center except for the case $ a=c=0 $. An additional random force $ \eta $ with zero mean and out of equilibrium
generates non-vanishing net currents and thus directed transport. In calculating these currents we pursue the approach followed by Machura et~al.\ in \cite{machura_2004} and restrict our analysis to dichotomous driving $ \eta \in \{ - \eta_0, + \eta_0 \}$ varying on a time scale much larger than all other characteristic time scales of the system (adiabatic limit).
 In this case, the behavior of the system may be described
 by stationary $L$-periodic solutions $ P_{\rm st, \pm\eta_0}(q) $ of the QSE corresponding to  constant currents $ J_{\pm\eta_0}= {\cal L} P_{\rm st, \pm\eta_0} $. If $ P_{\rm st, \pm\eta_0} $ is normalized
to one particle per unit cell, this current is for a given bias $ \eta $ equal to the mean velocity $\langle v\rangle_\eta= L J_\eta$ with
\begin{equation} \label{eqn:v}
\langle v \rangle_\eta =  \frac{1 - {\rm e}^{-\beta \eta}}{\int_0^L \! d q \, D_2 (q)^{-1} \, {\rm e}^{- \bar{\psi} (q) } \int_0^{L+q} \!\! d y \, {\rm e}^{+ \bar{\psi} (y) } d (y)^{-1} }\, .
\end{equation}
 Here, the coefficients $ d $ and $ D_2 $ are given by the respective orders in perturbation theory specified in Eqs.~(\ref{eqn:moyals0}), (\ref{eqn:moyals1}) or~(\ref{eqn:moyals2}) and the corresponding scaled minimal actions $\bar{\psi}=\psi/\hbar=\int_0^{q}dq' D_1(q)/D_2(q)$ read explicitly
\begin{align}
\label{actions_begin}
 \bar{\psi}^{(0)} (q) &= \beta V (q) - \beta \eta q - \frac{\beta^2}{2} \left( V' (q) - \eta \right)^2 \lambda \, , \\
 \bar{\psi}^{(1)} (q) & = \bar{\psi}^{(0)} (q) + \frac{\beta^3}{2} V'' (q) \left( V' (q) - \eta \right)^2 \lambda^2 \, , \\ \bar{\psi}^{(2)} (q) &= \bar{\psi}^{(1)} (q)\, .
\label{actions_end}
\end{align}
Note that the mean velocity (\ref{eqn:v}) does not depend on $ D_1 $ explicitly.\\
The net ratchet-current $ \bar{\langle v \rangle} = \left[ \langle v \rangle_{\eta_0} + \langle v \rangle_{- \eta_0} \right] / 2 $ is obtained by expanding
  $ \langle v \rangle_\eta $ to the respective order in $ \lambda $ and numerically evaluating the resulting integrals.

\begin{figure}
\includegraphics[width=8.5cm]{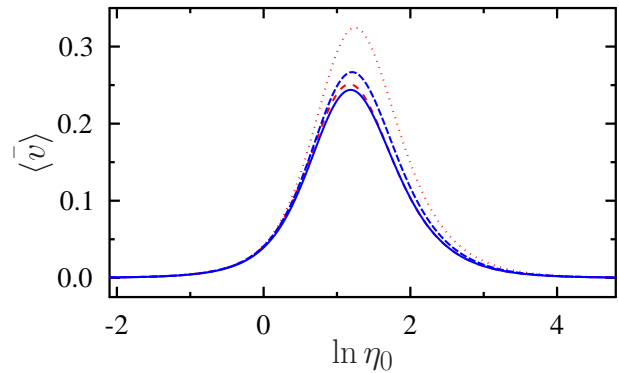}
 \caption{(Color online) Net ratchet current $ \bar{\langle v \rangle} $ vs noise amplitude $ \eta_0 $ for the potential (\ref{eqn:simple_pot}) for
$ \beta_0=5 $ and $ \lambda / L^2 = 0.0025 $. The classical result is represented by a dotted line, the curves including leading, locally harmonic and
higher-order quantum corrections are drawn as short dashed, long dashed and solid lines, respectively.}\label{fig:simple_ratchet}
\end{figure}
\begin{figure}
\includegraphics[width=8.5cm]{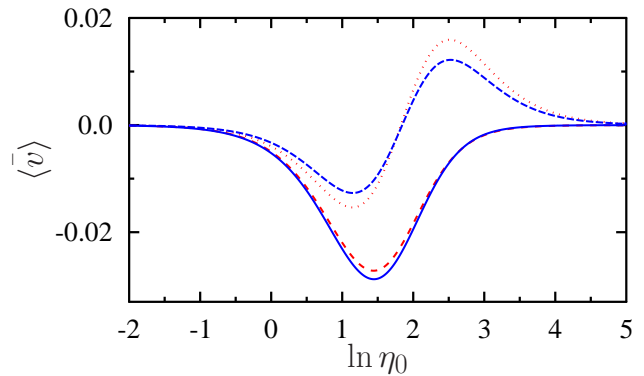}
 \caption{(Color online) Net ratchet current $ \bar{\langle v \rangle} $ vs noise amplitude $ \eta_0 $ for $ a = 0.4 $, $ b = 0.45 $, $ c =0.3 $,
$ \beta_0=2 $ and $ \lambda / L^2 = 0.0025 $. Line styles as in fig.~\ref{fig:simple_ratchet}.}\label{fig:rev_eta}
\end{figure}
\begin{figure}
\includegraphics[width=8.5cm]{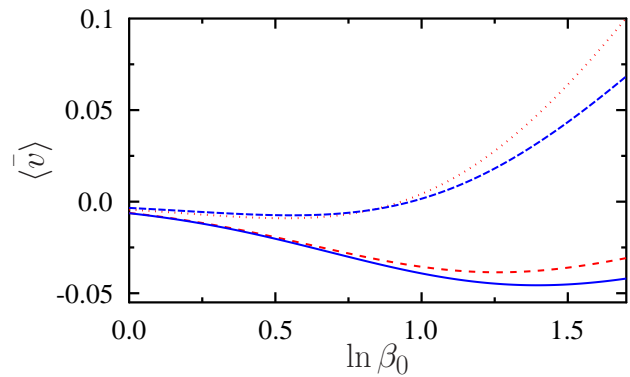}
 \caption{(Color online) Net ratchet current $ \bar{\langle v \rangle} $ vs inverse temperature $ \beta_0 $ for the same parameters as in fig.~\ref{fig:rev_eta}
 and $ \eta_0 = 5 $. Line styles as in fig.~\ref{fig:simple_ratchet}.}\label{fig:rev_beta}
\end{figure}

\section{Results and discussion}

In order to examine the influence of subleading quantum fluctuations based on a numerical evaluation of (\ref{eqn:v}), it is convenient to introduce  dimensionless quantities
 $ x= q/L $, $ \beta_0 = \beta \Delta V$, and $ W(x) = V(x) / \Delta V $ where the barrier height $ \Delta V $ is defined as the difference of the maximal and minimal value of the potential for vanishing bias.\\
We first look at an asymmetric potential with $b=c=0$ and the following parameter
\begin{equation} \label{eqn:simple_pot}
 W(x) = -0.454 \left[ \sin (2 \pi x ) + 0.25 \sin (4 \pi x ) \right] \, .
\end{equation}
 In fig.~\ref{fig:simple_ratchet}  the ratchet current is depicted for various orders of perturbation theory together with the classical result. Apparently, fluctuations always reduce the current compared to the classical case. However, the reduction due to the leading contributions (of order $\lambda$) is partially compensated and the net current pushed back towards its classical values when next order terms (local harmonic approximation) are considered as well. Only the inclusion of contributions of order $\lambda^2/L$ (local anharmonic approximation) induces smaller net currents again close to the results of the leading fluctuations. The convergence of the perturbation series is thus non-monotonous. The reason for this is a subtle interplay between higher-order contributions in the actions (\ref{actions_begin}-\ref{actions_end}) and the diffusion coefficients $D_2$. In local harmonic approximation one {\em always} has $D_2^{(1)}<D_2^{(0)}$ and the contribution of fluctuations in $\bar{\psi}^{(1)}$ is reduced by a factor $(1-\lambda\beta V'')$ compared to that in $\bar{\psi}^{(0)}$. In next order, while $\bar{\psi}^{(2)}=\bar{\psi}^{(1)}$, the diffusion coefficient $D_2^{(2)}$ is larger than $D_2^{(1)}$ in those regions of the ratchet potential, where $V'''<0$. Accordingly, terms produced in order $\lambda^2$ probe the asymmetry of the ratchet weaker than those in order $\lambda$, and thus soothe quantum effects in the net current. In terms of the Euclidian path integral for the equilibrium distribution, the local oscillator approximation
{\em symmetrizes} the potential experienced by the minimal-action path in a small vicinity around its starting and end point $ q $ with the tendency to suppress quantum phenomena due to asymmetries.

 Let us now analyze this scenario in a more involved case $ a = 0.4 $, $ b = 0.45 $, $ c =0.3 $, $ V_0 / \Delta V = 0.372 $.
 As already reported in \cite{machura_2004}, here leading quantum fluctuations generate a current reversal compared to the classical case, although the absolute value of quantum contributions to diffusion and action is small. This reveals the sensitivity of the current to even slight variations in the ratchet topology.
Results for increasing order in perturbation theory are shown in figs.~\ref{fig:rev_eta} and \ref{fig:rev_beta}. Notably, in this case data including second order harmonic terms (order $\lambda^2$) carry again the typical characteristics of the classical net current as a function of the noise amplitude $\eta_0$. Their impact is thus of the same size as that of the leading fluctuations. Only when also next order contributions induced by local asymmetries are taken into account, does one obtain results close to leading-order ones. The same applies for the situation in fig.~\ref{fig:rev_beta}, where the temperature dependence of the current is shown.
The conclusion is that in the sense of a consistent perturbative treatment, it is not justified to neglect formally smaller terms of order $\lambda^2/L$ against those of order $\lambda^2$.

\section{Summary}

In this paper we have analyzed the impact of subleading quantum corrections on net currents in overdamped ratchets at very low temperatures. A consistent inclusion of these contributions is non-trivial. Beyond the formal order of magnitude of the corresponding terms it has to take into account also their local symmetry properties in position space. Local anharmonicities are essential for a convergent perturbative treatment.
Along this rule, a systematic extension of the QSE provides a powerful tool to analyze dynamical properties of strongly condensed phase systems at low temperatures.

Financial support from the SFB569 is gratefully acknowledged.

\bibliography{biblio}

\end{document}